\documentclass[%
reprint,
superscriptaddress,
 amsmath,amssymb,
floatfix,
]{revtex4-1}

\usepackage[dvipdfmx]{graphicx}
\usepackage{dcolumn}
\usepackage{float}
\usepackage[dvipsnames]{xcolor}
\usepackage[breaklinks]{hyperref}

\begin{document}

\title{Feasibility of Kitaev quantum spin liquids in ultracold polar molecules}

\author{Kiyu Fukui}
\email{k.fukui@aion.t.u-tokyo.ac.jp}
\affiliation{Department of Applied Physics, The University of Tokyo, Bunkyo, Tokyo 113-8656, Japan}

\author{Yasuyuki Kato}%
\affiliation{Department of Applied Physics, The University of Tokyo, Bunkyo, Tokyo 113-8656, Japan}
 
 \author{Joji Nasu}
\affiliation{Department of Physics, Tohoku University, Sendai, Miyagi 980-8578, Japan}
\affiliation{PRESTO, Japan Science and Technology Agency, Honcho Kawaguchi, Saitama 332-0012, Japan}

\author{Yukitoshi Motome}
\affiliation{Department of Applied Physics, The University of Tokyo, Bunkyo, Tokyo 113-8656, Japan}

\date{\today}

\begin{abstract}
Ultracold atoms and molecules trapped in optical lattices are expected to serve as simulators of strongly correlated systems and topological states of matter. A fascinating example is to realize the Kitaev quantum spin liquid by using ultracold polar molecules. However, although experimental implementation of the Kitaev-type interaction was proposed, the stability of the Kitaev quantum spin liquid has not been fully investigated thus far. Here we study a quantum spin model with long-range angle-dependent Kitaev-type interactions proposed for the polar molecules, by the pseudofermion functional renormalization group method. We reveal that the ground state is magnetically ordered in both ferromagnetic and antiferromagnetic models regardless of the spatial anisotropy of the interactions, while the isotropic case is most frustrated and closest to the realization of the Kitaev quantum spin liquid. Furthermore, by introducing a cutoff in the interaction range, we clarify how the Kitaev quantum spin liquid is destroyed by the long-range interactions.  
The results urge us to reconsider the feasibility of the Kitaev quantum spin liquid in ultracold polar molecules.

\end{abstract}

\maketitle

\section{Introduction}\label{introduction}
The Kitaev model~\cite{Kitaev2006} provides us with a rare example of exact quantum spin liquid states~\cite{Diep, Balents2010, Lacroix, Zhou2017}, in more than one dimension. 
The model has bond-dependent anisotropic interactions on a honeycomb lattice, whose strong frustration results in a quantum spin liquid state with extremely short-range spin correlations~\cite{Baskaran2007}. In this quantum spin liquid state, which we call the Kitaev quantum spin liquid, the spins are fractionalized into itinerant Majorana fermions and localized $\mathbb{Z}_2$ gauge fluxes. This provides a good playground for the Majorana fermions, which have been explored for many years in particle physics~\cite{Majorana1937}. Careful comparison between experimental data and theoretical results has accumulated evidence of such exotic quasiparticle excitations. These fractional excitations are expected to be utilized for topological quantum computation, and have been attracting great interest from a wide range of fields, including not only condensed matter physics but also quantum information~\cite{Kitaev2003, Kitaev2006}. 

The search for candidate materials for the Kitaev model has been actively conducted since its realization mechanism was proposed for strongly correlated electron systems~\cite{Jackeli2009}. A number of candidates, called Kitaev materials, have been revealed by the intensive search from both experimental and theoretical perspectives~\cite{rau2016, Trebst2017, Winter2017, Takagi2019, Motome2020a, Motome2020, Trebst2022}, for example, Na$_2$IrO$_3$~\cite{Chaloupka2010, Singh2010, Singh2012, Comin2012, Foyevtsova2013, Sohn2013, Katukuri2014, Yamaji2014, HwanChun2015, Winter2016}, $\alpha$-Li$_2$IrO$_3$~\cite{Singh2012, Winter2016}, and $\alpha$-RuCl$_3$~\cite{Plumb2014, Kubota2015, Winter2016, Yadav2016, Sinn2016}. However, due to competing magnetic interactions that inevitably appear in the solid state realizations, such as the Heisenberg exchange interaction, almost all the candidates undergo a phase transition to a magnetically ordered phase at low temperatures. 
Hence, it is still a challenging task to materialize the pristine Kitaev spin liquid.

A different realization of the Kitaev spin liquid has been proposed in ultracold polar molecules trapped in optical lattices. Ultracold atoms and molecules are known to provide a platform for studying strongly correlated systems~\cite{Lewenstein2007, Bloch2008, Gadway2016, Zhang2018}. 
Among them, ultracold polar molecules, such as KRb~\cite{Ni2008} and LiCs~\cite{Deiglmayr2008, Deiglmayr2009}, trapped in optical lattices have been expected to serve as good simulators of quantum magnets~\cite{Barnett2006, Micheli2006}. In particular, experimentally accessible implementation of various spin lattice models with arbitrary spin lengths $S\geq1/2$ was theoretically proposed by using microwave dressed states of molecules~\cite{Manmana2013}. In this context, a possible realization of the Kitaev-type interaction was also proposed~\cite{Manmana2013, Gorshkov2013}. In this proposal, the bond-dependent anisotropic interactions are mimicked by angle-dependent dipole interactions between molecules. However, the interactions are long ranged with a spatial decay of $r^{-3}$, where $r$ is the distance between the molecules, it is left as an open question whether the Kitaev quantum spin liquid can survive against such long-range interactions. 

In this paper, we present our numerical results on the ground state of a spin model with long-range dipolar interactions proposed in the previous studies, which we call the dipolar Kitaev model, by using the pseudofermion functional renormalization group (PFFRG) method. The PFFRG is a powerful numerical method which is capable of dealing with a wide range of the spin models even in the presence of strong frustration and long-range interactions~\cite{Reuther2009, Reuther2010}. Calculating the spin susceptibility by the PFFRG, we clarify that the frustration of the dipolar Kitaev model is much weaker than that of the original Kitaev model, and the ground state is always magnetically ordered regardless of the spatial anisotropy of the interactions in both ferromagnetic (FM) and antiferromagnetic (AFM) cases. We also unravel how the Kitaev quantum spin liquid becomes unstable against the long-range interactions while changing the range of the interactions.

The structure of this paper is as follows. In Sec.~\ref{model}, we introduce the dipolar Kitaev model. In Sec.~\ref{method}, we briefly review the PFFRG method and present the conditions of our numerical calculations. We present our results on the dipolar Kitaev model for the FM and AFM cases in Secs.~\ref{subsec:FM} and \ref{subsec:AFM}, respectively. In addition, we analyze the effect of anisotropy in the interactions in Sec.~\ref{subsec:frustration} and the effect of long-range interactions in Sec.~\ref{subsec:long-range}. In Sec.~\ref{discussion}, we discuss our results. Finally, we summarize our main findings in Sec.~\ref{conclusion}.

\section{Model}\label{model}
\begin{figure}
    \centering
    \includegraphics[width=0.9\columnwidth, clip]{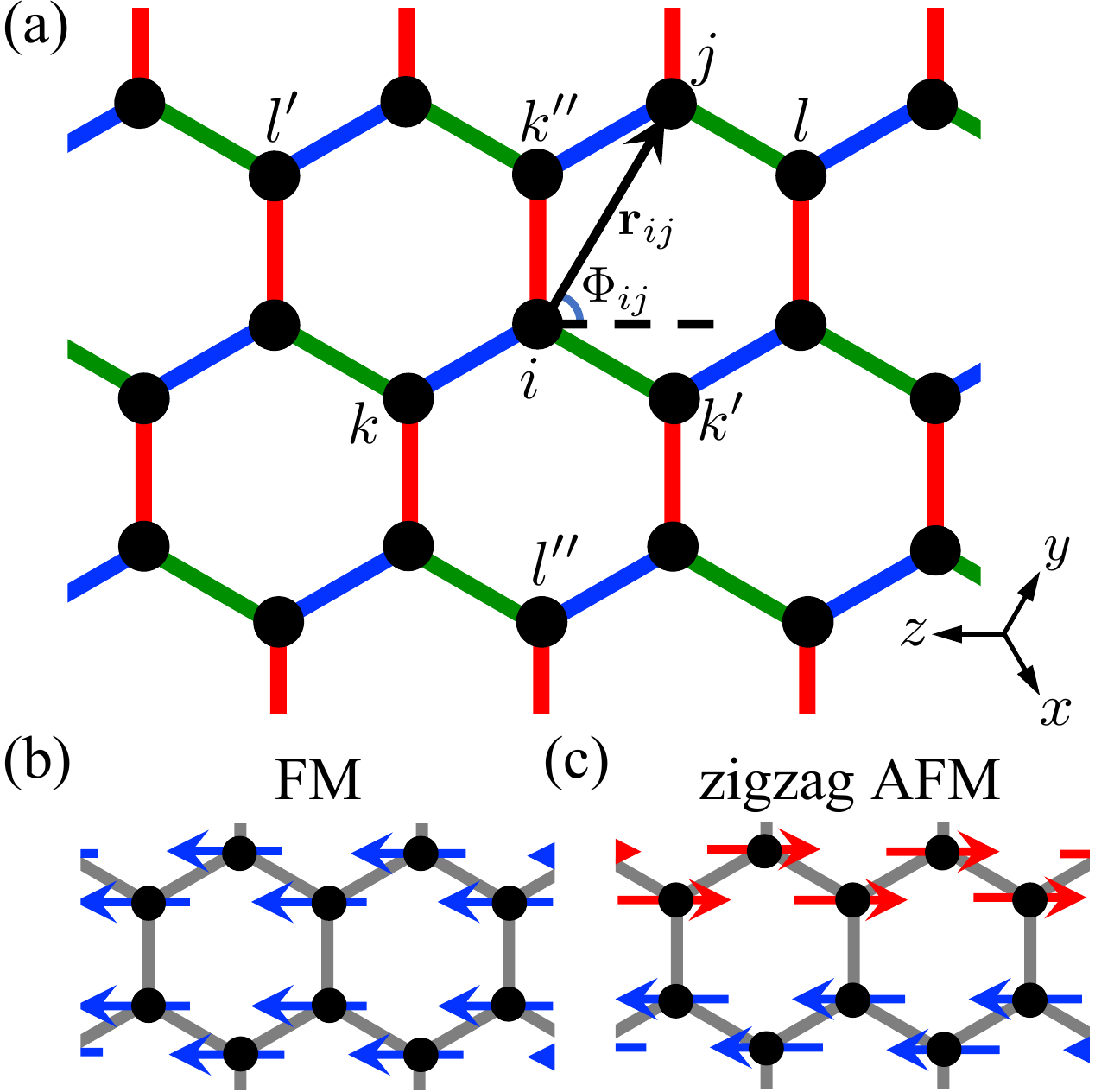}
    \caption{(a) Schematic figure for the dipolar Kitaev model in Eq.~\eqref{dipolar_Kitaev_model}. The blue, green, and red bonds represent the nearest-neighbor couplings $J_x$, $J_y$, and $J_z$, respectively. The lower right arrows indicate spin axes. (b) and (c) Spin configurations in the ferromagnetic (FM) and zigzag antiferromagnetic (AFM) states, which are realized in the FM and AFM dipolar Kitaev models, respectively. The spin directions depend on $J_{x}$, $J_{y}$, and $J_{z}$; the cases with $J_{z}>J_{x}$ and $J_{y}$ are shown.}
    \label{fig:lattice_dipolar_Kitaev}
\end{figure}

 Following the previous studies~\cite{Manmana2013, Gorshkov2013}, we introduce a model for implementation of the Kitaev-type interaction in ultracold polar molecules trapped in an optical honeycomb lattice, which we call the dipolar Kitaev model. The Hamiltonian is given by 
\begin{align}\label{dipolar_Kitaev_model}
    \mathcal{H}=\sum_{i<j}H_{ij} 
\;\;\;\;&\notag\\
=\sum_{i<j}\frac{-1}{3r_{ij}^3}\biggl\{&J_x\left[1-2\cos\left(2\Phi_{ij}-\frac{4\pi}{3}\right)\right]S_i^xS_j^x\notag\\
+&J_y\left[1-2\cos\biggl(2\Phi_{ij}-\frac{2\pi}{3}\biggr)\right]S_i^yS_j^y\notag\\
+&J_z\left[1-2\cos\big(2\Phi_{ij}\big)\right]S_i^zS_j^z\biggr\}, 
\end{align}
where $r_{ij}=\lvert\mathbf{r}_{ij}\rvert$
is the distance between sites $i$ and $j$ on a honeycomb lattice (we set the length of the nearest-neighbor bond as unity), and $\Phi_{ij}$ is the angle of the direction from $i$ to $j$, as shown in Fig.~\ref{fig:lattice_dipolar_Kitaev}(a); $J_{\mu}$ is the coupling constant and $S_i^{\mu}$ represents the $\mu$ component of the $S=1/2$ quantum spin at site $i$ ($\mu=x$, $y$, and $z$). Although only the isotropic FM case where $J_x=J_y=J_z>0$ was considered in the previous studies~\cite{Manmana2013, Gorshkov2013}, we extend the model to anisotropic cases where $J_\mu$ are not equivalent including the AFM case, as the interactions are expected to be controlled in a wide range by microwave irradiation~\cite{Gorshkov2013}. 

The model in Eq.~\eqref{dipolar_Kitaev_model} has interactions depending on the angle between two spins. This is a generalization of the bond-dependent nearest-neighbor couplings in the original Kitaev model~\cite{Kitaev2006} to the long-range dipolar form, as explained below. 
For a nearest-neighbor bond, say a blue bond in Fig.~\ref{fig:lattice_dipolar_Kitaev}(a), $r_{ik}=1$ and $\Phi_{ik}=-\frac{5\pi}{6}$, and hence, $H_{ij}$ in Eq.~\eqref{dipolar_Kitaev_model} becomes $H_{ik}=-J_xS_i^xS_{k}^x$. In a similar manner, we obtain $H_{ik'}=-J_yS_i^yS_{k'}^y$ and $H_{ik''}=-J_zS_i^zS_{k''}^z$ on the green and red bond with $\Phi_{ik'}=-\frac{\pi}{6}$ and $\Phi_{ik''}=\frac{\pi}{2}$, respectively. Therefore, the interactions between nearest-neighbor spins are the same with those in the Kitaev model~\cite{Kitaev2006}.
On the other hand, for a third-neighbor bond, for example, between the sites $i$ and $l$ in Fig.~\ref{fig:lattice_dipolar_Kitaev}(a), which is parallel to the nearest-neighbor blue bond, $r_{il}=2$ and $\Phi_{il}=\frac{\pi}{6}$, and therefore, $H_{ij}$ in Eq.~\eqref{dipolar_Kitaev_model} becomes $H_{il}=-\frac{J_x}{8}S_i^xS_{l}^x$. In the similar manner, we obtain $H_{il'}=-\frac{J_y}{8}S_i^yS_{l'}^y$ and $H_{il''}=-\frac{J_z}{8}S_i^zS_{l''}^z$ for third-neighbor bonds between $i$ and $l'$, and $i$ and $l''$, respectively, in Fig.~\ref{fig:lattice_dipolar_Kitaev}(a). 
These have the same bond-dependent form as the nearest-neighbor ones besides the coefficient of $1/8$ from the decay factor $1/r_{ij}^3$ similar to the conventional dipolar interaction. 
Meanwhile, for a second-neighbor bond, for instance, between the sites $i$ and $j$
in Fig.~\ref{fig:lattice_dipolar_Kitaev}(a), $r_{ij}=\sqrt{3}$ and $\Phi_{ij}=\frac{\pi}{3}$, 
and hence, we obtain 
$H_{ij}=-\frac{1}{9\sqrt{3}}\left[2J_xS_i^xS_{j}^x-J_yS_i^yS_{j}^y+2J_zS_i^zS_{j}^z\right]$.
Thus, in general, all the diagonal components of two-spin interactions appear with amplitudes and signs depending on the angle between the two spins. Note that these interactions for the second-neighbor bonds are different from those discussed in the previous studies~\cite{Reuther2014a, Rousochatzakis2015}.

\section{method}\label{method}
We study the ground state of the dipolar Kitaev model in Eq.~\eqref{dipolar_Kitaev_model}, by using the PFFRG method. The PFFRG is a powerful numerical method for quantum spin systems~\cite{Reuther2009, Reuther2010}, which has been successfully applied to a number of 2D and 3D models for frustrated quantum magnets with Heisenberg interactions~\cite{Reuther2009, Reuther2010}, $XXZ$ interactions~\cite{Gottel2012, Buessen2018}, Kitaev-like interactions~\cite{Reuther2011a, Reuther2012, Reuther2014a, Revelli2019a}, and nondiagonal interactions~\cite{Hering2017, Buessen2019}. 
Furthermore, the extensions to systems with $S>1/2$~\cite{Baez2017, Buessen2018, Iqbal2019} or SU($N$)~\cite{Buessen2018a, Roscher2018, Roscher2019} quantum spins were proposed. It was also applied to the dipolar Heisenberg model with long-range interactions~\cite{Keles2018, Keles2018a}. 

In the PFFRG method for the $S=1/2$ systems, the spin operator is written in terms of auxiliary fermions~\cite{Abrikosov1965}, called pseudofermions, as
\begin{align}\label{pseudofermion}
    S^{\mu}_i=\frac{1}{2}\sum_{\alpha, \alpha'}f^\dagger_{i\alpha'}\sigma^{\mu}_{\alpha',
\alpha}f_{i\alpha},
\end{align}
where $f_{i\alpha}$ ($f^\dagger_{i\alpha})$ is an annihilation (creation) operator of the pseudofermion at site $i$ with spin $\alpha\in\{\uparrow, \downarrow\}$, and $\sigma^{\mu}$ is the $\mu$ component of the Pauli matrices (we set the reduced Planck constant $\hbar$ as unity). As this spin-fermion mapping enlarges the Hilbert space, a pure-imaginary chemical potential $\mu=-\frac{\mathrm{i}\pi}{2\beta}$, where $\beta$ is inverse temperature, is often introduced to restrict the Hilbert space to the local half-filled subspace with $\sum_{\alpha}f^\dagger_{i\alpha}f_{i\alpha}=1$~\cite{Popov1988}. In the present study, however, we do not need such a prescription since we focus on the zero-temperature limit ($\beta \to \infty$) where the local constraint is fulfilled automatically. By using Eq.~\eqref{pseudofermion}, the bilinear spin Hamiltonian in Eq.~\eqref{dipolar_Kitaev_model} is rewritten into a quartic one in terms of the pseudofermions.
In the following, we adopt the fermionic one-particle irreducible FRG~\cite{Salmhofer, Salmhofer2001, Kopietz, Metzner2012, Platt2013} for the quartic fermionic Hamiltonian.

The PFFRG is performed by the fermionic FRG flow equations for the self-energy and two-particle vertex function. We employ one-loop truncation in a fully self-consistent form~\cite{Katanin2004}, in which the fermionic FRG flow equations for the self-energy $\Sigma$ and the two-particle vertex function $\Gamma$ are given by~\cite{Salmhofer2001, Kopietz, Metzner2012, Platt2013}
\begin{align}
    \frac{\mathrm{d}}{\mathrm{d}\Lambda}\Sigma^{\Lambda}_{x_1'; x_1}=&\sum_{x_2, x_2'}\Gamma^{\Lambda}_{x_1', x_2'; x_1, x_2}\mathcal{S}^{\Lambda}_{x_2, x_2'},\label{eq:self_energy_flow}\\
    \frac{\mathrm{d}}{\mathrm{d}\Lambda}\Gamma^{\Lambda}_{x_1', x_2'; x_1, x_2}=&-\sum_{x_3, x_4, x_3', x_4'}L^{\Lambda}_{x_3, x_4; x_3', x_4'}\notag\\
    &\times\biggl[\frac{1}{2}\Gamma^{\Lambda}_{x_1', x_2'; x_3, x_4}\Gamma^{\Lambda}_{x_3', x_4; x_1, x_2}\notag\\
    &\ \ \ \ \ -\Gamma^{\Lambda}_{x_1', x_4'; x_1, x_3}\Gamma^{\Lambda}_{x_3', x_2'; x_4, x_2}\notag\\
    &\ \ \ \ \ +\Gamma^{\Lambda}_{x_2', x_4'; x_1, x_3}\Gamma^{\Lambda}_{x_3', x_1'; x_4, x_2}\biggr],\label{eq:vertex_flow}
\end{align}
respectively, where $\Lambda$ is the energy cutoff scale in the renormalization group method. 
Here, $x=(\omega,\ i,\ \alpha)$ denotes a set of the Matsubara frequency $\omega$, the lattice site $i$, and the spin index $\alpha$, for which the summation $\sum_{x}$ is taken as $\int^{\infty}_{-\infty}\frac{\mathrm{d}\omega}{2\pi}\sum_{i}\sum_{\alpha}$, since the Matsubara frequency is a continuous variable in the zero-temperature limit. 
In Eq.~\eqref{eq:self_energy_flow}, $\mathcal{S}^{\Lambda}$ is the single-scale propagator regularized by the cutoff scale $\Lambda$, which is defined by the bare propagator $G^{\Lambda}_{0}$ and the full propagator $G^{\Lambda}$ as 
\begin{align}\label{eq:def_S}
    \mathcal{S}^{\Lambda}_{x;x'}=-
\left[G^{\Lambda}\frac{\mathrm{d}(G^{\Lambda}_{0})^{-1}}{\mathrm{d}\Lambda}G^{\Lambda}\right]_{x;x'},
\end{align}
with
\begin{align}
G^{\Lambda}_{0, x;x'}&=2\pi\delta(\omega-\omega')\delta_{i, i'}\delta_{\alpha,\alpha'}\frac{\Theta(\lvert \omega\rvert-\Lambda)}{\mathrm{i}\omega},\\
   G^{\Lambda}_{x;x'}&=-\int^{\beta}_{0}\mathrm{d}\tau\mathrm{d}\tau'\  \mathrm{e}^{\mathrm{i}(\omega\tau-\omega'\tau')}\langle T_{\tau}f_{i\alpha}(\tau)f^{\dagger}_{i'\alpha'}(\tau')\rangle_{\Lambda},
\end{align}
where $\langle T_{\tau}\cdots\rangle_{\Lambda}$ means the expectation value of the imaginary-time-ordered operators $f_{i\alpha}(\tau)=\mathrm{e}^{\tau\mathcal{H}}f_{i\alpha}\mathrm{e}^{-\tau\mathcal{H}}$ and $f^\dagger_{i'\alpha'}(\tau')=\mathrm{e}^{\tau'\mathcal{H}}f^\dagger_{i'\alpha'}\mathrm{e}^{-\tau'\mathcal{H}}$ with the cutoff energy scale $\Lambda$. 
Here, $\Theta(x)$ is the Heaviside function, which works as the cutoff function for the FRG to project out all the modes for $\lvert\omega\rvert<\Lambda$, and $\delta(x)$ is the delta function. 

$G^\Lambda$ and $G_0^\Lambda$ are related with $\Sigma^\Lambda$ as
\begin{align}
    G^{\Lambda}_{x;x'}=\left[(G^{\Lambda}_0)^{-1}-\Sigma^{\Lambda}\right]^{-1}_{x;x'}.
\end{align}
Meanwhile, $L^{\Lambda}$ in Eq.~\eqref{eq:vertex_flow} is defined as
\begin{equation}\label{eq:def_L}
    L^{\Lambda}_{x_1,x_2;x_1',x_2'}=\frac{\mathrm{d}}{\mathrm{d}\Lambda}\left(G^{\Lambda}_{x_1;x_1'}G^{\Lambda}_{x_2;x_2'}\right).
\end{equation}
We porperly handle ambiguity arising from the derivatives of the Heaviside function in Eqs.~\eqref{eq:def_S} and~\eqref{eq:def_L}~\cite{MORRIS1994}.

In the following, we study magnetic instabilities in the paramagnetic state where all the lattice sites are equivalent. In the present system, the pseudofermions are localized at each site due to the lack of the bilinear kinetic energy term. Using
this locality and the energy conservation law, we can parametrize the self-energy, the full propagator, and the single-scale propagator as~\cite{Reuther2010, Reuther}
\begin{align}
    \mathcal{O}^{\Lambda}_{x';x}=2\pi\delta(\omega-\omega')\delta_{i,i'}\delta_{\alpha,\alpha'}\mathcal{O}^{\Lambda}(\omega),
\end{align}
for $\mathcal{O}=\Sigma$, $G$, and $\mathcal{S}$, where
\begin{align}\label{eq:G_basis}
    G^{\Lambda}(\omega)
=\frac{\Theta(\lvert\omega\rvert-\Lambda)}{\mathrm{i}\omega-\Sigma^{\Lambda}(\omega)},
\quad
    \mathcal{S}^{\Lambda}(\omega)
=-\frac{\delta(\lvert\omega\rvert-\Lambda)}{\mathrm{i}\omega-\Sigma^{\Lambda}(\omega)}.
\end{align}
Meanwhile, under the locality of the pseudofermions, the two-particle vertex function depends only on two site indices as
\begin{align}\label{eq:Gamma_site_basis}
    \Gamma^{\Lambda}_{x_1',x_2';x_1,x_2}&=
\Gamma^{\Lambda}_{i_1i_2}(\omega_1'\alpha_1',\omega_2'\alpha_2'; \omega_1\alpha_1, \omega_2\alpha_2)\delta_{i_1',i_1}\delta_{i_2',i_2}\notag\\
    &-\Gamma^{\Lambda}_{i_1i_2}(\omega_2'\alpha_2',\omega_1'\alpha_1'; \omega_1\alpha_1, \omega_2\alpha_2)\delta_{i_2',i_1}\delta_{i_1',i_2},
\end{align}
which can be parametrized by using
\begin{align}
    \Gamma^{\Lambda}_{i_1i_2}&(\omega_1'\alpha_1',\omega_2'\alpha_2';\omega_1\alpha_1,\omega_2\alpha_2)\notag\\
    &=2\pi\delta(\omega_1'+\omega_2'-\omega_1-\omega_2)\notag\\
    &\times\biggl\{\sum_{\mu=x,y,z}\Gamma^{\mu,\Lambda}_{i_1i_2}(s,t,u)\sigma^{\mu}_{\alpha_1',\alpha_1}\sigma^{\mu}_{\alpha_2',\alpha_2}\notag\\
    &\ \ \ \ \ \ \ \ +\Gamma^{\mathrm{d},\Lambda}_{i_1i_2}(s,t,u)\delta_{\alpha_1',\alpha_1}\delta_{\alpha_2',\alpha_2}\biggr\}\label{eq:Gamma_basis},
\end{align}
with
\begin{align}
    s=\omega_1'+\omega_2',
\
    t=\omega_1'-\omega_1,
\
    u=\omega_1'-\omega_2.
\end{align}
Here, $s$, $t$, and $u$ correspond to the transfer energies in the particle-particle, direct particle-hole, and crossed particle-hole scattering channels, respectively~\cite{Metzner2012,Reuther}. 
In Eq.~\eqref{eq:Gamma_basis}, $\Gamma^{\mu,\Lambda}_{i_1i_2}(s, t, u)$ represents the renormalized dynamical coupling between the $\mu$ component of the pseudofermion spins, while $\Gamma^{\mathrm{d},\Lambda}_{i_1i_2}(s, t, u)$ represents the density-density coupling between pseudofermions which is generated through the renormalization process. Note that the parametrization in Eq.~\eqref{eq:Gamma_basis} is applicable to the diagonal interactions with anisotropy like the Kitaev model; more general expression for nondiagonal interactions is found in Ref.~\cite{Buessen2019}.

The fully parametrized flow equations are obtained by substituting Eqs.~\eqref{eq:G_basis}, \eqref{eq:Gamma_site_basis}, and \eqref{eq:Gamma_basis} into Eqs.~\eqref{eq:self_energy_flow} and \eqref{eq:vertex_flow}~\cite{Reuther2010, Reuther, Gottel, Baez, Hering, Buessen, Fukui}. To solve the integro-differential equations in Eqs.~\eqref{eq:self_energy_flow} and~\eqref{eq:vertex_flow}, we start from the initial conditions given by
\begin{align}
    \Sigma^{\Lambda\to\infty}(\omega)&=0,\\
    \Gamma^{x,\Lambda\to\infty}_{i_1i_2}(s,t,u)&=
\frac{-J_x}{3r_{i_1i_2}^3}\left[1-2\cos
\left(2\Phi_{i_1i_2}-\frac{4\pi}{3}
\right)\right],\\
    \Gamma^{y,\Lambda\to\infty}_{i_1i_2}(s,t,u)&=
\frac{-J_y}{3r_{i_1i_2}^3}\left[1-2\cos
\left(2\Phi_{i_1i_2}-\frac{2\pi}{3}
\right)\right],\\
    \Gamma^{z,\Lambda\to\infty}_{i_1i_2}(s,t,u)&=
\frac{-J_z}{3r_{i_1i_2}^3}\left[1-2\cos\left(2\Phi_{i_1i_2}\right)\right],\\
    \Gamma^{\mathrm{d},\Lambda\to\infty}_{i_1i_2}(s,t,u)&=0.
\end{align}
After solving the FRG flow equations, we calculate observables of the original spin systems from the obtained self-energy and vertex function. In the following calculations, to detect magnetic instabilities, we compute the diagonal components of the spin susceptibility as
\begin{widetext}
\begin{align}
 \chi^{\mu\mu, \Lambda}_{ij}&=\int^{\infty}_0\mathrm{d}\tau\ \langle T_{\tau}S^{\mu}_i(\tau)S^{\mu}_j(0)\rangle_{\Lambda}\notag\\ 
    &=-\int^{\infty}_{-\infty}\frac{\mathrm{d}\omega}{4\pi}\ G^{\Lambda}(\omega)^2\delta_{i,j}-\int^{\infty}_{-\infty}\frac{\mathrm{d}\omega\mathrm{d}\omega'}{8\pi^2}\ G^{\Lambda}(\omega)^2G^{\Lambda}(\omega')^2\big[2\Gamma^{\mu, \Lambda}_{ij}(\omega+\omega', 0, \omega-\omega')-\big\{\Gamma^{\mu,\Lambda}_{ii}(\omega+\omega', \omega-\omega', 0)\notag\\
    &\qquad\qquad\qquad\qquad\qquad\qquad\qquad\qquad\qquad\qquad\qquad\; -\sum_{\nu\neq\mu}\Gamma^{\nu, \Lambda}_{ii}(\omega+\omega', \omega-\omega', 0)+\Gamma^{\mathrm{d}, \Lambda}_{ii}(\omega+\omega', \omega-\omega', 0)\big\}\delta_{i,j}],
\end{align}
\end{widetext}
using the self-energy and vertex function with the cutoff scale $\Lambda$ obtained from the renormalization group calculation. 
When the system shows an instability toward a magnetically ordered state, one can detect it from the $\Lambda$ dependence of the spin susceptibility; it is signaled by the divergence of the Fourier transform $\chi^{\mu\mu, \Lambda}(\mathbf{k})$ at momentum $\mathbf{k}$ corresponding to the ordering vector, where $\chi^{\mu\mu, \Lambda}(\mathbf{k})=\frac{1}{N}\sum_{i, j}\chi^{\mu\mu, \Lambda}_{ij}\mathrm{e}^{\mathrm{i}\mathbf{k}\cdot\mathbf{r}_{ij}}$ with the number of sites $N$. 
We call this critical value of $\Lambda$ the critical cutoff scale $\Lambda_{\mathrm{c}}$.
In practice, however, due to the finite system size and the finite frequency grid, the $\Lambda$ dependence of $\chi^{\mu\mu, \Lambda}(\mathbf{k})$ shows a kink or cusp instead of the divergence. 
Hence, we use such an anomaly to detect the magnetic instability and estimate $\Lambda_{\mathrm{c}}$. 
On the other hand, when $\chi^{\mu\mu, \Lambda}(\mathbf{k})$ changes smoothly at all $\mathbf{k}$ down to $\Lambda\rightarrow 0$, the system does not undergo any instability, suggesting the realization of a quantum spin liquid state in the ground state.

In the following numerical calculations, we use the logarithmic frequency grid with 64 positive frequency points between 10$^{-4}$ and 250. We also generate the logarithmic $\Lambda$ grid starting from $\Lambda_{\mathrm{max}}=500$ to $\Lambda_{\mathrm{min}}\simeq10^{-2}$ by multiplying a factor of $0.95$. In the calculations, we neglect two-particle vertex functions between two sites further apart than $L=20$ lattice, which corresponds to a finite-size cluster containing $N=631$ lattice sites. 
We show the dependences of $\chi^{zz,\Lambda}(\mathbf{k})$ on the number of $\omega$ and $\Lambda$ grids in Appendix~\ref{appx:omega_Lambda} and on the system size $L$ in Appendix~\ref{appx:scaling}.
Furthermore, we present the finite-size scaling of $\chi^{zz,\Lambda}(\mathbf{k})$ to estimate $\Lambda_{\mathrm{c}}$ in the thermodynamic limit of $L\to \infty$, and conclude that the typical error of $\Lambda_{\mathrm{c}}$ calculated for $L=20$ is roughly $10$~\% in Appendix~\ref{appx:scaling}.

\section{Result}\label{result}
\subsection{Ferromagnetic case}\label{subsec:FM}
First, we study the ground state of the dipolar Kitaev model in Eq.~\eqref{dipolar_Kitaev_model} for the FM case with $J_x\geq 0$, $J_y\geq 0$, and $J_z\geq 0$.
Assuming $J_x=J_y$, we parametrize the anisotropy as
\begin{align}\label{eq:FM_param}
    J_x=J_y=\alpha,
\quad
    J_z=3-2\alpha,
\end{align}
where $0\leq \alpha\leq 1.5$. The isotropic case of $J_x=J_y=J_z$ corresponds to $\alpha=1$. 

\begin{figure}
    \centering
    \includegraphics[width=0.86\columnwidth, clip]{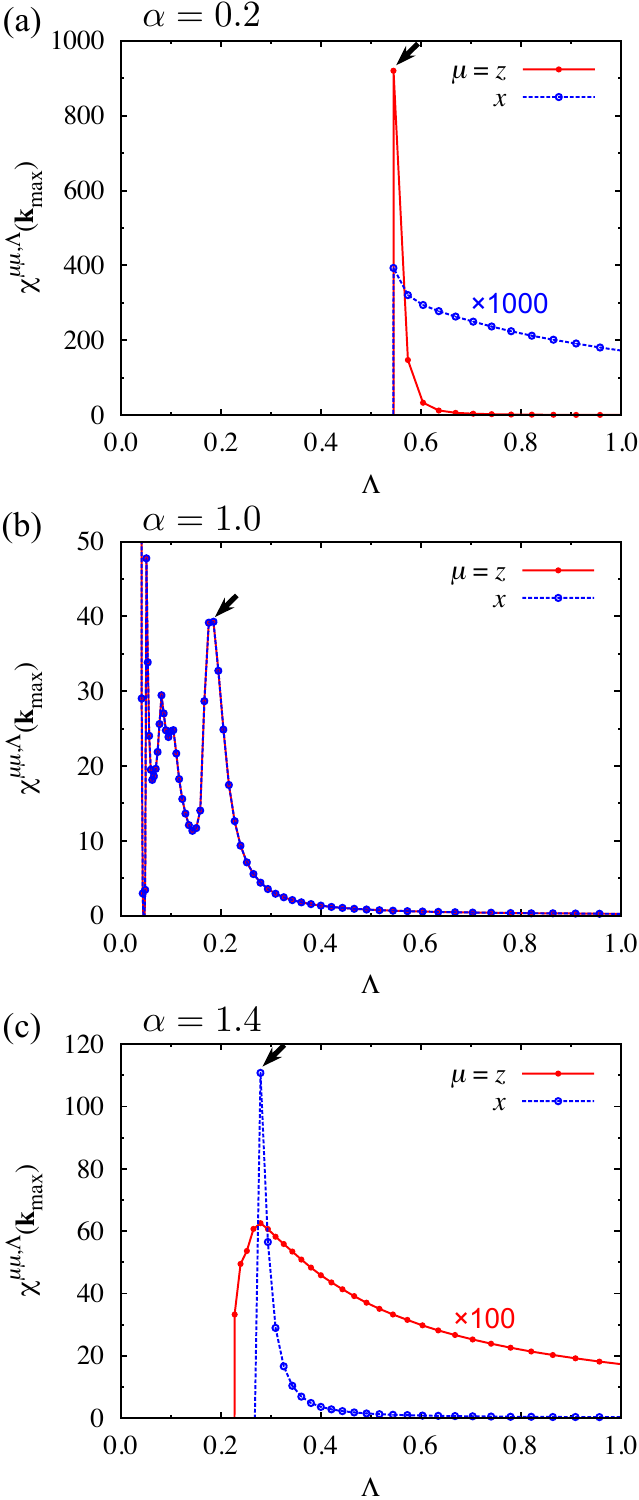}
    \caption{Spin susceptibilities $\chi^{zz,\Lambda}(\bf{k}_{\mathrm{max}})$ and $\chi^{xx,\Lambda}(\bf{k}_{\mathrm{max}})$ as functions of the cutoff scale $\Lambda$ for the FM dipolar Kitaev model with (a) $\alpha=0.2$, (b) $\alpha=1.0$, and (c) $\alpha=1.4$. $\bf{k}_{\mathrm{max}}$ is the wave vector at which the susceptibility becomes maximum in the reciprocal space; in this FM case, $\bf{k}_{\mathrm{max}}$ is always located at $\bf{k}_{\mathrm{max}}=0$; see Fig.~\ref{fig:FM_suscep_map}. The black arrows indicate the critical cutoff scale $\Lambda_{\mathrm{c}}$.}
    \label{fig:FM_suscep_flow}
\end{figure}

\begin{figure}
    \centering
    \includegraphics[width=1.0\columnwidth, clip]{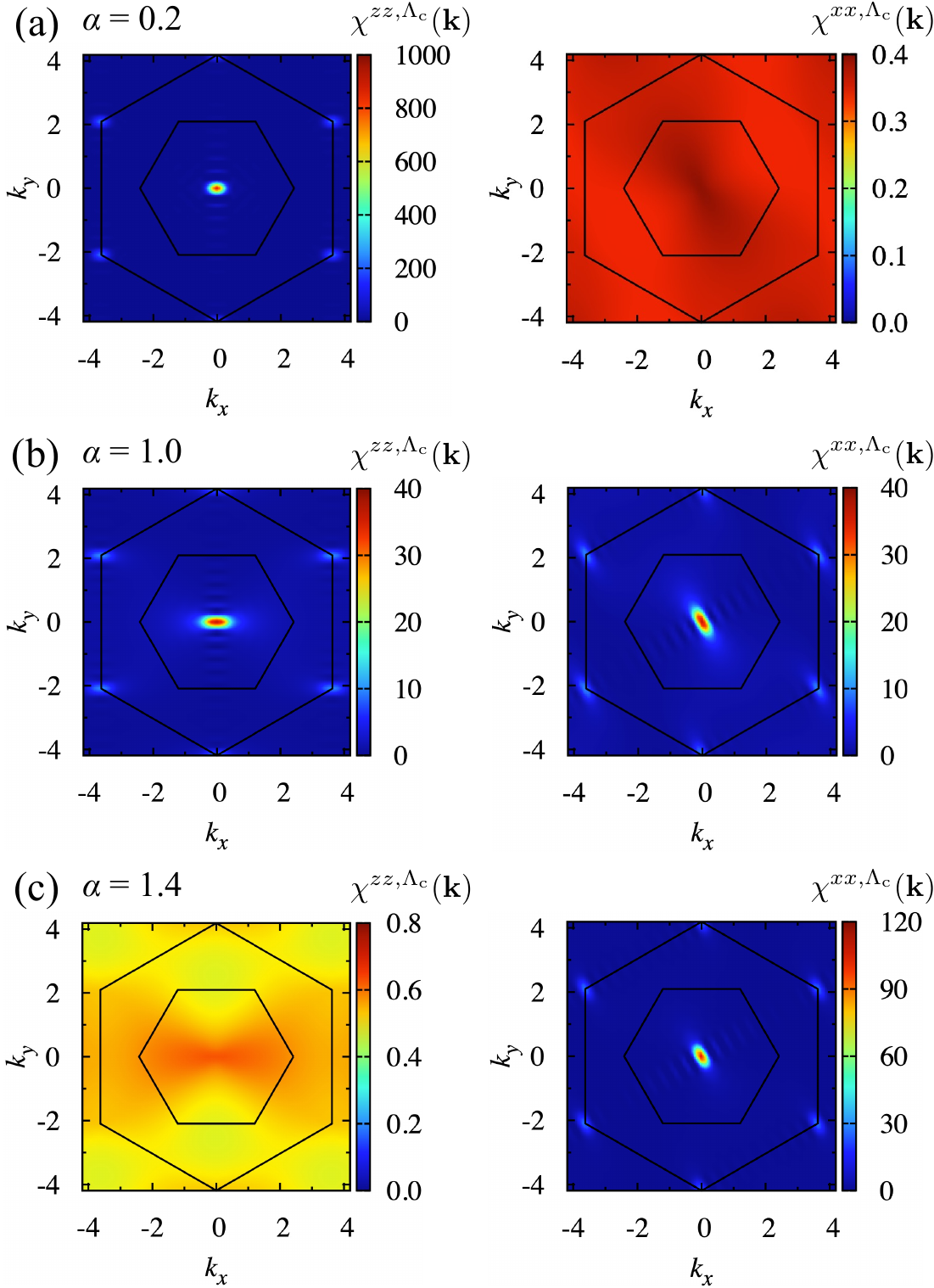}
    \caption{Contour plots of $\chi^{zz,\Lambda}(\bf{k})$ (left) and $\chi^{xx,\Lambda}(\bf{k})$ (right) at $\Lambda=\Lambda_{\mathrm{c}}$ in the reciprocal space of $\mathbf{k}=(k_x,k_y)$ for the FM dipolar Kitaev model with (a) $\alpha=0.2$, (b) $\alpha=1.0$, and (c) $\alpha=1.4$. The inner black hexagon indicates the first Brillouin zone, while the outer one indicates the zone including up to the third Brillouin zones.}
    \label{fig:FM_suscep_map}
\end{figure}

Figure~\ref{fig:FM_suscep_flow} shows the $\Lambda$ dependences of $\chi^{zz,\Lambda}(\mathbf{k}_{\mathrm{max}})$ and $\chi^{xx,\Lambda}(\mathbf{k}_{\mathrm{max}}) = \chi^{yy,\Lambda}(\mathbf{k}_{\mathrm{max}})$, where $\mathbf{k}_{\mathrm{max}}$ represents the wave vector at which the susceptibility takes a maximum in the reciprocal space, for three values of $\alpha$: (a) $\alpha=0.2$ ($J_z=2.6 > J_{x}=J_y=0.2$), (b) $\alpha=1.0$ ($J_x=J_y=J_z=1.0$), and (c) $\alpha=1.4$ ($J_x=J_y=1.4 > J_z=0.2$). 
For all the cases, we find that the susceptibility shows a maximum at $\mathbf{k}_{\mathrm{max}} = \bf{0}$ as shown in Fig.~\ref{fig:FM_suscep_map}, indicating that FM spin fluctuations are dominant.
$\chi^{zz,\Lambda}(\mathbf{k})$ is always larger (smaller) than $\chi^{xx,\Lambda}(\mathbf{k})$ for $0\leq \alpha<1.0$ ($1.0<\alpha\leq 1.5$), while $\chi^{xx,\Lambda}(\mathbf{k}) = \chi^{yy,\Lambda}(\mathbf{k}) = \chi^{zz,\Lambda}(\mathbf{k})$ for the isotropic case of $\alpha=1.0$. 
As shown in Fig.~\ref{fig:FM_suscep_flow}, $\chi^{zz,\Lambda}(\mathbf{k}_{\mathrm{max}}=\mathbf{0})$ and $\chi^{xx,\Lambda}(\mathbf{k}_{\mathrm{max}}=\mathbf{0})$ show sharp peaks or cusps at some value of $\Lambda$, as indicated by the black arrows in the figures. 
These indicate magnetic instabilities toward the FM ordered state. 
Similar instabilities are found for other values of $\alpha$; see Sec.~\ref{subsec:frustration}. 
Thus, we conclude that the ground state of the FM dipolar Kitaev model is in the FM ordered phase regardless of the value of $\alpha$. 
A schematic figure for the FM ordered state is shown in Fig.~\ref{fig:lattice_dipolar_Kitaev}(b).
The values of the critical cutoff scale $\Lambda_{\mathrm{c}}$ depend on $\alpha$, and takes a minimum around the isotropic case of $\alpha=1.0$, as discussed in Sec.~\ref{subsec:frustration}.

\subsection{Antiferromagnetic case}\label{subsec:AFM}
\begin{figure}
    \centering
    \includegraphics[width=0.86\columnwidth, clip]{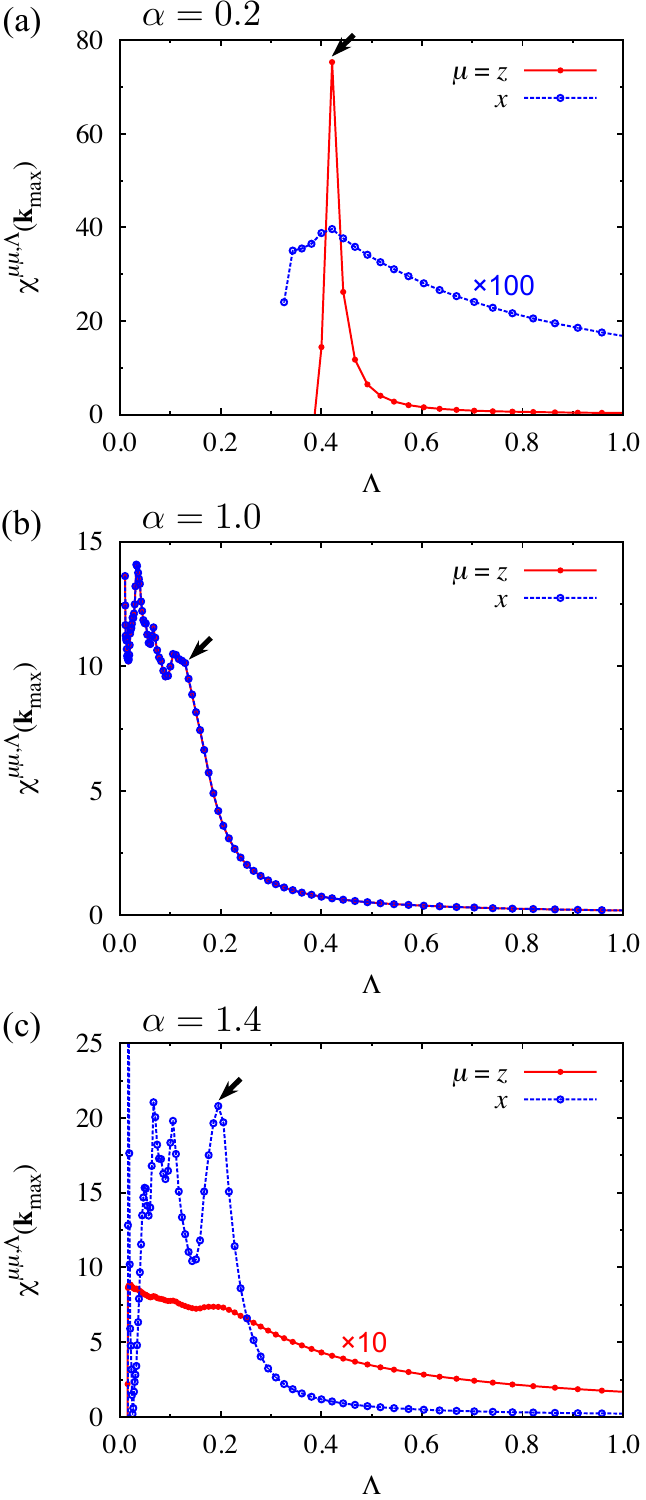}
    \caption{Spin susceptibilities $\chi^{zz,\Lambda}(\bf{k}_{\mathrm{max}})$ and $\chi^{xx,\Lambda}(\bf{k}_{\mathrm{max}})$ as functions of $\Lambda$ for the AFM dipolar Kitaev model with (a) $\alpha=0.2$, (b) $\alpha=1.0$, and (c) $\alpha=1.4$. In this AFM case, $\bf{k}_{\mathrm{max}}$ is always located at $(0, \pm\frac{2\pi}{3})$ and $(\pm\frac{\pi}{\sqrt{3}},\pm\frac{\pi}{3})$ for $\chi^{zz,\Lambda}(\bf{k})$ and $\chi^{xx,\Lambda}(\bf{k})$, respectively; see Fig.~\ref{fig:AFM_suscep_map}. The notations are common to those in Fig.~\ref{fig:FM_suscep_flow}.}
    \label{fig:AFM_suscep_flow}
\end{figure}

\begin{figure}
    \centering
    \includegraphics[width=1.0\columnwidth, clip]{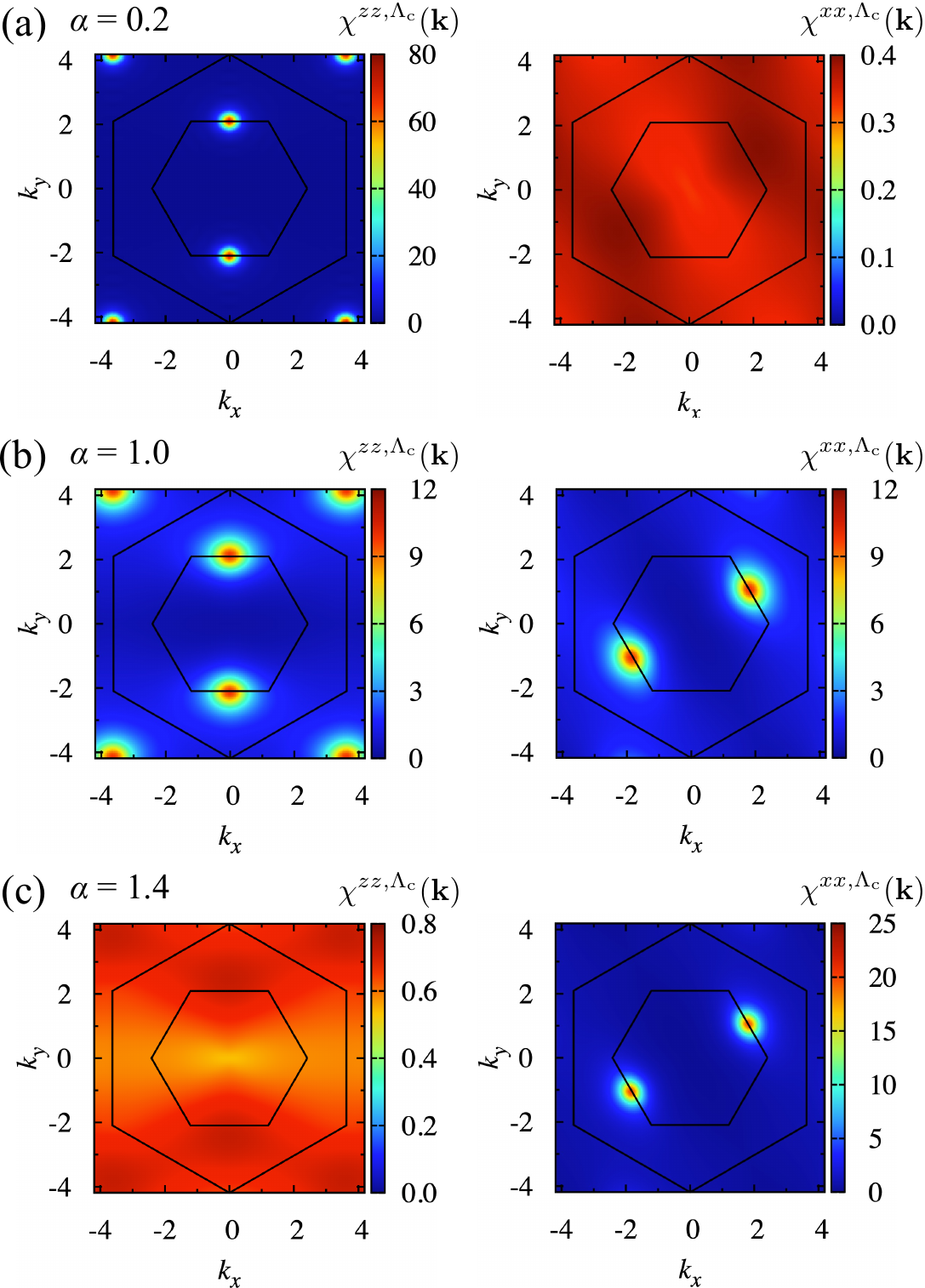}
    \caption{Contour plots of $\chi^{zz,\Lambda_{\mathrm{c}}}(\bf{k})$ and $\chi^{xx,\Lambda_{\mathrm{c}}}(\bf{k})$ for the AFM dipolar Kitaev model with (a) $\alpha=0.2$, (b) $\alpha=1.0$, and (c) $\alpha=1.4$. The notations are common to those in Fig.~\ref{fig:FM_suscep_map}.}
    \label{fig:AFM_suscep_map}
\end{figure}

Then, let us move on to the AFM case with $J_x\leq0$, $J_y\leq0$, and $J_z\leq0$. In a similar manner to the FM case, we assume $J_x=J_y$ and parametrize the anisotropy as
\begin{align}
    J_x=J_y=-\alpha,
\quad
    J_z=-(3-2\alpha), 
\end{align}
where $0\leq \alpha\leq 1.5$. Figure~\ref{fig:AFM_suscep_flow} shows the $\Lambda$ dependences of $\chi^{zz,\Lambda}(\mathbf{k}_{\mathrm{max}})$ and $\chi^{xx,\Lambda}(\mathbf{k}_{\mathrm{max}})$, for three values of $\alpha$: (a) $\alpha=0.2$ ($J_z=-2.6 < J_x=J_y=-0.2$), (b) $\alpha=1.0$ ($J_x=J_y=J_z=-1.0$), and (c) $\alpha=1.4$ ($J_x=J_y=-1.4 < J_z=-0.2$). 
Note that the relation $\chi^{xx,\Lambda}(k_x, k_y) = \chi^{yy,\Lambda}(-k_x, k_y)$ holds in the AFM case. For all $\alpha$, we find that $\chi^{zz,\Lambda}(\mathbf{k})$ and $\chi^{xx,\Lambda}(\mathbf{k})$ show maxima at $\mathbf{k}_{\mathrm{max}} = (0, \pm\frac{2\pi}{3})$ and  $(\pm\frac{\pi}{\sqrt{3}},\pm\frac{\pi}{3})$, respectively, as shown in Fig.~\ref{fig:AFM_suscep_map}. 
Similar to the FM case, we find anomalies at some value of $\Lambda$, which in this case indicates magnetic instabilities toward the zigzag AFM ordered state characterized by the peaks at $\mathbf{k}_{\mathrm{max}}$.
The spin configuration in this state is shown in Fig.~\ref{fig:lattice_dipolar_Kitaev}(c).
Thus, we conclude that the ground state of the AFM dipolar Kitaev model is in the zigzag AFM ordered phase regardless of $\alpha$. 
As in the FM case, the values of $\Lambda_{\mathrm{c}}$ depend on $\alpha$, and takes a minimum around the isotropic case of $\alpha=1.0$; we will discuss this behavior in Sec.~\ref{subsec:frustration}.

Let us discuss why the zigzag state is preferred in the ground state in the AFM case, instead of a simple collinear N\'eel state. 
The spin configuration in the zigzag state has energy gain from all second- and third-neighbor bonds, in addition to one of the nearest-neighbor bonds; see Fig.~\ref{fig:lattice_dipolar_Kitaev}(c) and the forms of the interactions discussed in Sec.~\ref{model}.
In contrast, the N\'eel AFM state does not have energy gain from the second neighbors where the spins are ferromagnetically aligned, although it gains energy from the nearest and third neighbors similar to the zigzag case. Thus, this simple consideration implies that the Kitaev-type second-neighbor interactions play an important role for the formation of the zigzag AFM order rather than the N\'eel one. However, we will show that to stabilize the zigzag AFM long-range order, the second-neighbor interactions are not sufficient and that further-neighbor ones are necessary in Sec.~\ref{subsec:long-range}.


\subsection{$\alpha$ dependence of $\Lambda_{\mathrm{c}}$}\label{subsec:frustration}
\begin{figure}
    \centering
    \includegraphics[width=1.0\columnwidth, clip]{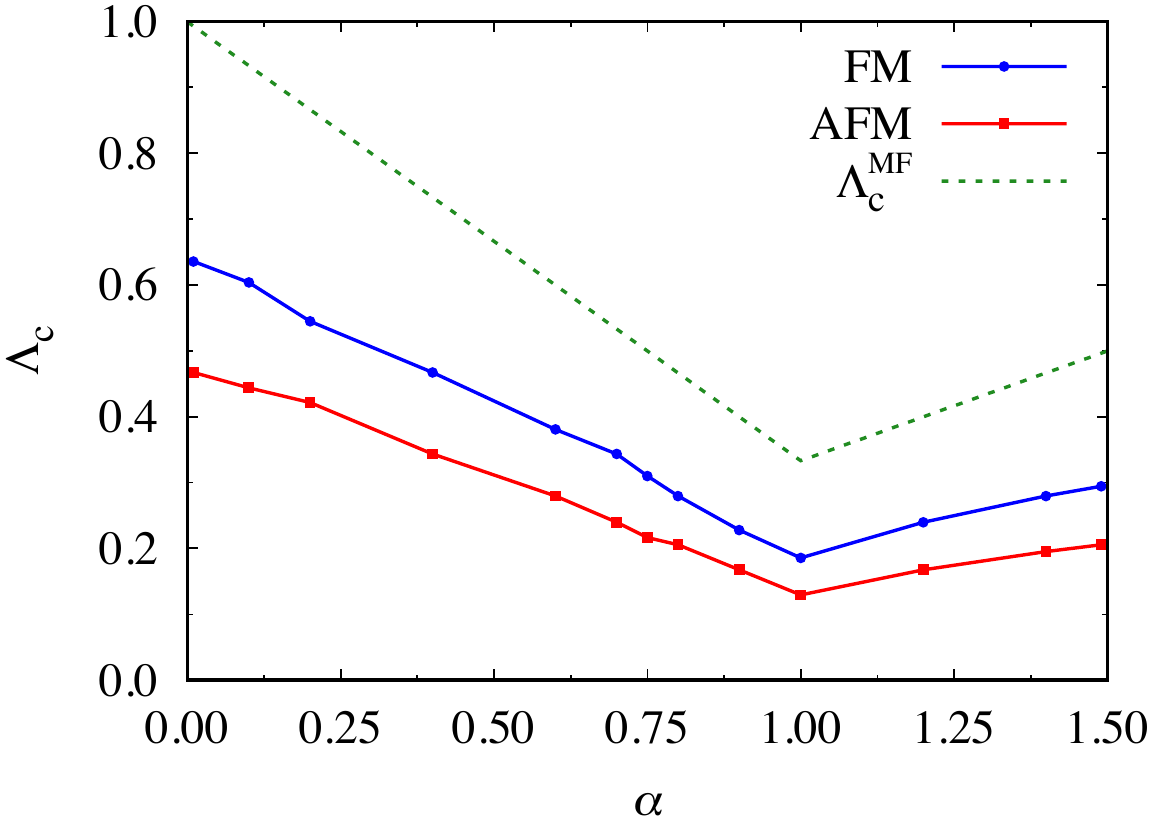}
    \caption{$\alpha$ dependences of $\Lambda_{\mathrm{c}}$ and $\Lambda_{\rm c}^{\rm MF}$ to the FM and AFM dipolar Kitaev models. $\Lambda_{\rm c}^{\rm MF}$ is common to both FM and AFM cases.}
    \label{fig:FM_AFM_summary}
\end{figure}

As shown in Secs.~\ref{subsec:FM} and \ref{subsec:AFM}, the ground states for the FM and AFM dipolar Kitaev models are the FM and zigzag AFM ordered states, respectively. 
We obtain the same conclusions for other values of $\alpha$ as well as for several parameters in the fully anisotropic cases where all $J_\mu$ are inequivalent.  
Figure~\ref{fig:FM_AFM_summary} summarizes the $\alpha$ dependences of $\Lambda_{\mathrm{c}}$ for both FM and AFM cases. We find that $\Lambda_{\mathrm{c}}$ becomes smallest for the isotropic cases at $\alpha=1.0$ for both FM and AFM cases; it increases almost linearly while both decreasing and increasing $\alpha$. We note that $\Lambda_{\mathrm{c}}$ for AFM case is lower than that for FM case for all $\alpha$.
The results indicate that the dipolar Kitaev model does not realize a quantum spin liquid state for all $\alpha$, while the minimum at $\alpha=1.0$  suggests that the isotropic case is closest to its realization. In other words, the frustration of the dipolar Kitaev model becomes strongest for the isotropic coupling constants $J_x=J_y=J_z$ in both FM and AFM cases, where the system is closest to the realization of quantum spin liquid. 

In Fig.~\ref{fig:FM_AFM_summary}, we also plot a mean-field estimate of $\Lambda_{\rm c}$, $\Lambda_{\rm c}^{\rm MF}=\frac{2}{\pi}\lvert\Theta_{\mathrm{CW}}\rvert$, where $\Theta_{\mathrm{CW}}$ is the Curie-Weiss temperature and the factor of $\frac{2}{\pi}$ comes from the approximate relation $\Lambda\simeq\frac{2}{\pi}T$ between $\Lambda$ and temperature $T$~\cite{Iqbal2016, Buessen2016}. Here, $\Theta_{\mathrm{CW}}$ is obtained as $\Theta_{\mathrm{CW}}=-\frac{1}{4}\sum_{j}\mathcal{J}_{ij}$, where $\mathcal{J}_{ij}$ is the coefficient of $S^{z}_{i}S^{z}_{j}$ ($S^{x}_{i}S^{x}_{j}$ or $S^{y}_{i}S^{y}_{j}$) in Eq.~\eqref{dipolar_Kitaev_model} for $0\leq\alpha\leq1.0$ ($1.0\leq\alpha\leq1.5$), and the summation $\sum_{j}$ is taken for the cluster used in the PFFRG calculations with the central site $i$. We find that $\Lambda_{\rm c}^{\rm MF}$ shows similar $\alpha$ dependence to $\Lambda_{\mathrm{c}}$: it depends linearly with respect to $\alpha$ and becomes smallest at $\alpha=1.0$. The values of $\Lambda_{\rm c}^{\rm MF}$ are about $1.6$ ($2.1$) times larger than $\Lambda_{\rm c}$ for the FM (AFM) case. The results suggest that while $\Lambda_{\rm c}^{\rm MF}$ describes the overall behavior of $\Lambda_{\rm c}$, the magnetic instability is suppressed by quantum fluctuations beyond the mean-field approximation.

\subsection{Effect of long-range dipolar interactions}\label{subsec:long-range}
Since the Kitaev model, which has nearest-neighbor interactions only, is known to give a quantum spin liquid ground state~\cite{Kitaev2006}, our results indicate that the long-range dipolar interactions hamper its realization and cause the instabilities toward magnetic orderings. 
To elucidate the effect of the long-range interactions, here we vary the range of the interactions by introducing a cutoff length $L_{\mathrm{int}}$ for the model in Eq.~\eqref{dipolar_Kitaev_model}; namely, we take the summation of $i$ and $j$ in Eq.~\eqref{dipolar_Kitaev_model} only within the range of $\|\mathbf{r}_{ij}\|_{\mathrm{b}}\leq L_{\mathrm{int}}$, where
$\|\mathbf{r}_{ij}\|_{\mathrm{b}}$ is the bond distance between sites $i$ and $j$ on the honeycomb lattice (for instance, $\|\mathbf{r}_{ij}\|_{\mathrm{b}}=2$ for second-neighbor sites). 
Then, the model with $L_{\mathrm{int}}=1$ is equivalent to the original Kitaev model, while that with $L_{\mathrm{int}}\to \infty$ corresponds to the dipolar Kitaev model in Eq.~\eqref{dipolar_Kitaev_model}.
In the following, we study the ground state while changing $L_{\mathrm{int}}$ for the isotropic case of $\alpha=1.0$ in the FM model.

\begin{figure}
    \centering
    \includegraphics[width=1.0\columnwidth, clip]{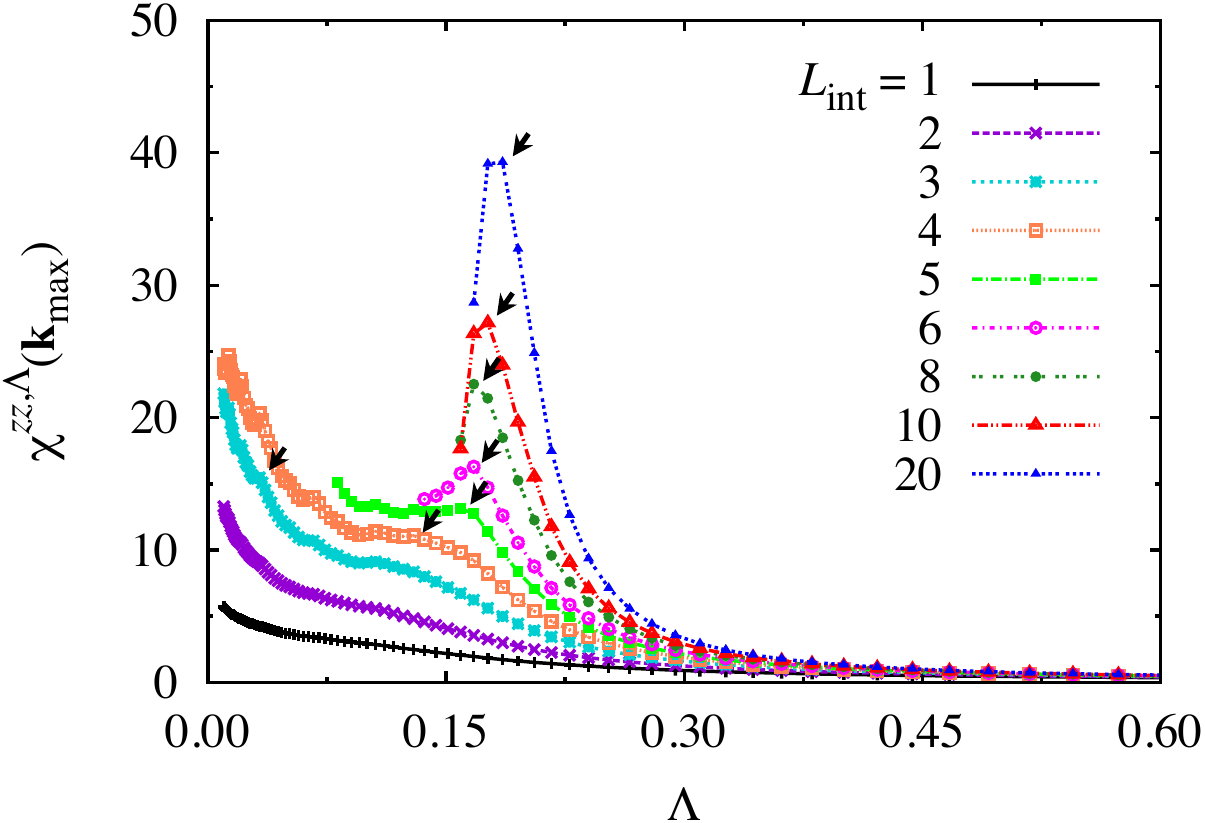}
    \caption{Spin susceptibility $\chi^{zz,\Lambda}(\bf{k}_{\mathrm{max}})$ of the isotropic FM dipolar Kitaev systems ($\alpha=1$) for several values of the cutoff in the interaction range, $L_{\mathrm{int}}$. The black arrows indicate $\Lambda_{\mathrm{c}}$. See the text and Fig.~\ref{fig:Lint_suscep_map} for the values of $\mathbf{k}_{\mathrm{max}}$.}
    \label{fig:Lint_suscep_flow}
\end{figure}

Figure~\ref{fig:Lint_suscep_flow} shows the $\Lambda$ dependences of $\chi^{zz,\Lambda}(\mathbf{k}_{\mathrm{max}})$ while changing $L_{\mathrm{int}}$ from $1$ to $20$.
We find that $\chi^{zz,\Lambda}(\mathbf{k}_{\mathrm{max}})$ shows no apparent anomaly down to the smallest $\Lambda$ when $L_{\mathrm{int}}=1$ and $2$, whereas it shows a peak or kink for larger $L_{\mathrm{int}}$ at $\Lambda_{\mathrm{c}}$ indicated by the black arrows in Fig.~\ref{fig:Lint_suscep_flow}. Although the locations of $\Lambda_{\mathrm{c}}$ are subtle for $L_{\mathrm{int}}=3$ and $4$, we also carefully examine the system size dependence of the local susceptibility to identify $\Lambda_{\mathrm{c}}$, following the previous studies~\cite{Kiese2019, Buessen2021}; the details are described in Appendix~\ref{appx:local_suscep}.
Thus, our PFFRG results indicate that (i) the Kitaev quantum spin liquid is obtained at $L_{\mathrm{int}}=1$ consistent with the exact solution, (ii) it appears to survive for $L_{\mathrm{int}}=2$, but (iii) it is replaced by the ordered state for $L_{\mathrm{int}}\geq3$.
In other words, our results indicate that the third-neighbor interactions are sufficient to kill the Kitaev quantum spin liquid.

\begin{figure}
    \centering
    \includegraphics[width=1.0\columnwidth, clip]{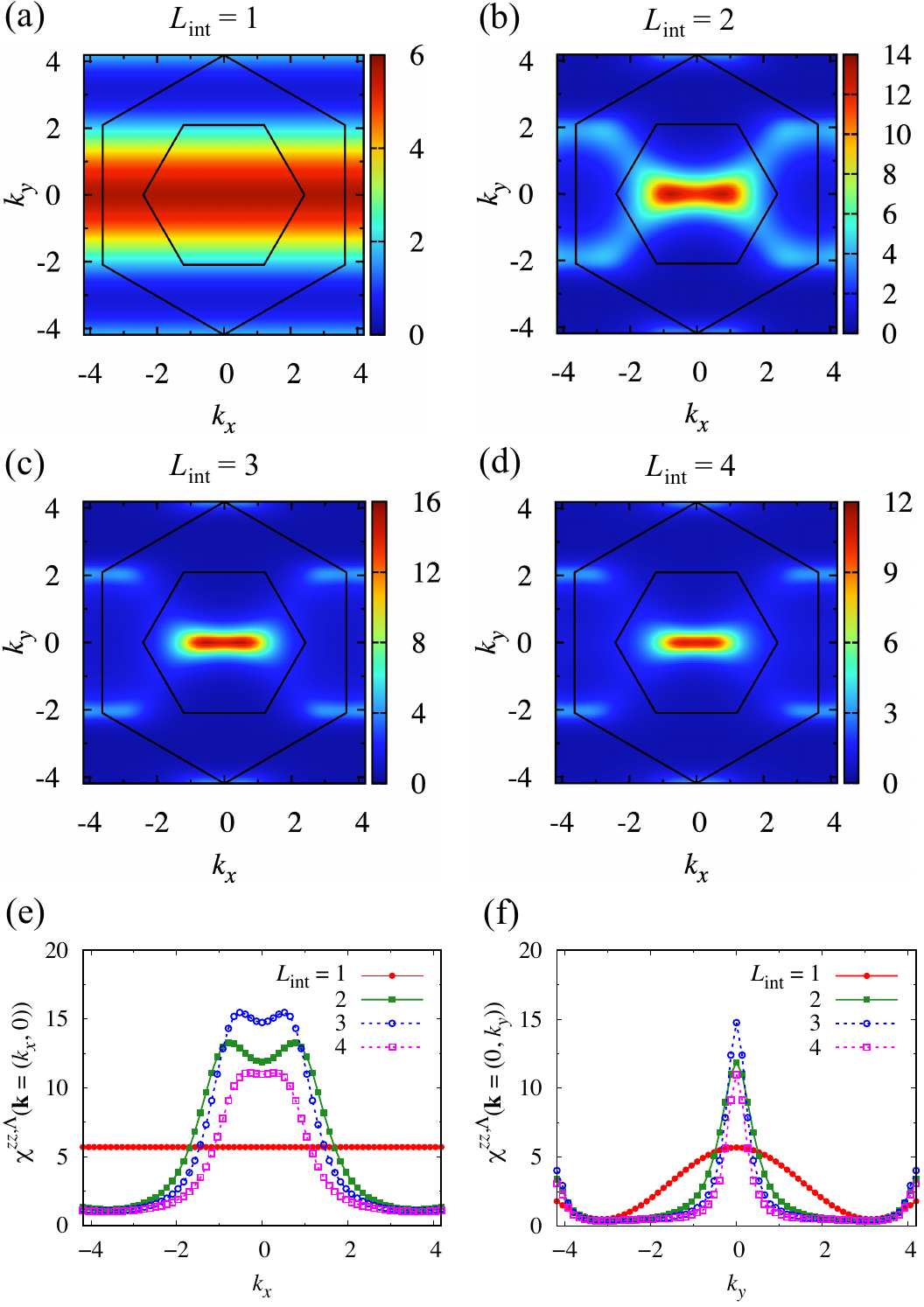}
    \caption{Contour plots of $\chi^{zz,\Lambda}(\bf{k})$ for the isotropic FM dipolar Kitaev model with (a) $L_{\mathrm{int}}=1$, (b) $L_{\mathrm{int}}=2$,  (c) $L_{\mathrm{int}}=3$, and (d) $L_{\mathrm{int}}=4$. 
    (e) and (f) Profiles of (a)--(d) at $k_y=0$ and $k_x=0$, respectively. The data for $L_{\mathrm{int}}=1$ and $2$ in (a), (b), (e), and (f) are at the minimum cutoff scale $\Lambda_{\mathrm{min}}$, while those for $L_{\mathrm{int}}=3$ and $4$ in (c), (d), (e), and (f) are at the critical cutoff scale $\Lambda_{\mathrm{c}}$.}
    \label{fig:Lint_suscep_map}
\end{figure}

We show the $\mathbf{k}$ dependence of $\chi^{zz,\Lambda}(\mathbf{k})$ for $L_{\mathrm{int}}=1$, $2$, $3$, and $4$ in Fig.~\ref{fig:Lint_suscep_map}.
The data for $L_{\mathrm{int}}=1$ and $2$ are obtained at the smallest value of $\Lambda=\Lambda_{\mathrm{min}}$, while those for $L_{\mathrm{int}}=3$ and $4$ are at $\Lambda=\Lambda_{\mathrm{c}}$. When $L_{\mathrm{int}}=1$ for which the system corresponds to the original Kitaev model, $\chi^{zz,\Lambda}(\mathbf{k})$ is proportional to $\cos k_y$ as shown in Figs.~\ref{fig:Lint_suscep_map}(a) and \ref{fig:Lint_suscep_map}(f), while it does not depend on $k_{x}$ as shown in Fig.~\ref{fig:Lint_suscep_map}(e), reflecting the fact that the spin correlations are nonzero only for nearest-neighbor spins~\cite{Baskaran2007}. While increasing $L_{\mathrm{int}}$, the broad peak of the cosine curve at $\mathbf{k}=(0, 0)$ shrinks in both $k_x$ and $k_y$ directions and grows into a sharp peak with strong intensity as shown in Figs.~\ref{fig:Lint_suscep_map}(b)--\ref{fig:Lint_suscep_map}(f), corresponding to the FM ordering.
Note that when $L_{\mathrm{int}}$ is small, $\chi^{zz, \Lambda}(\mathbf{k})$ shows peaks at $\mathbf{k}_{\mathrm{max}}=(k_{x}\neq 0, k_{y}=0)$; the peaks approach $\mathbf{k}=\mathbf{0}$ while increasing $L_{\mathrm{int}}$, and $\mathbf{k}_{\mathrm{max}}$ becomes zero for $L_{\mathrm{int}} \gtrsim 12$.

\begin{figure}
    \centering
    \includegraphics[width=1.0\columnwidth, clip]{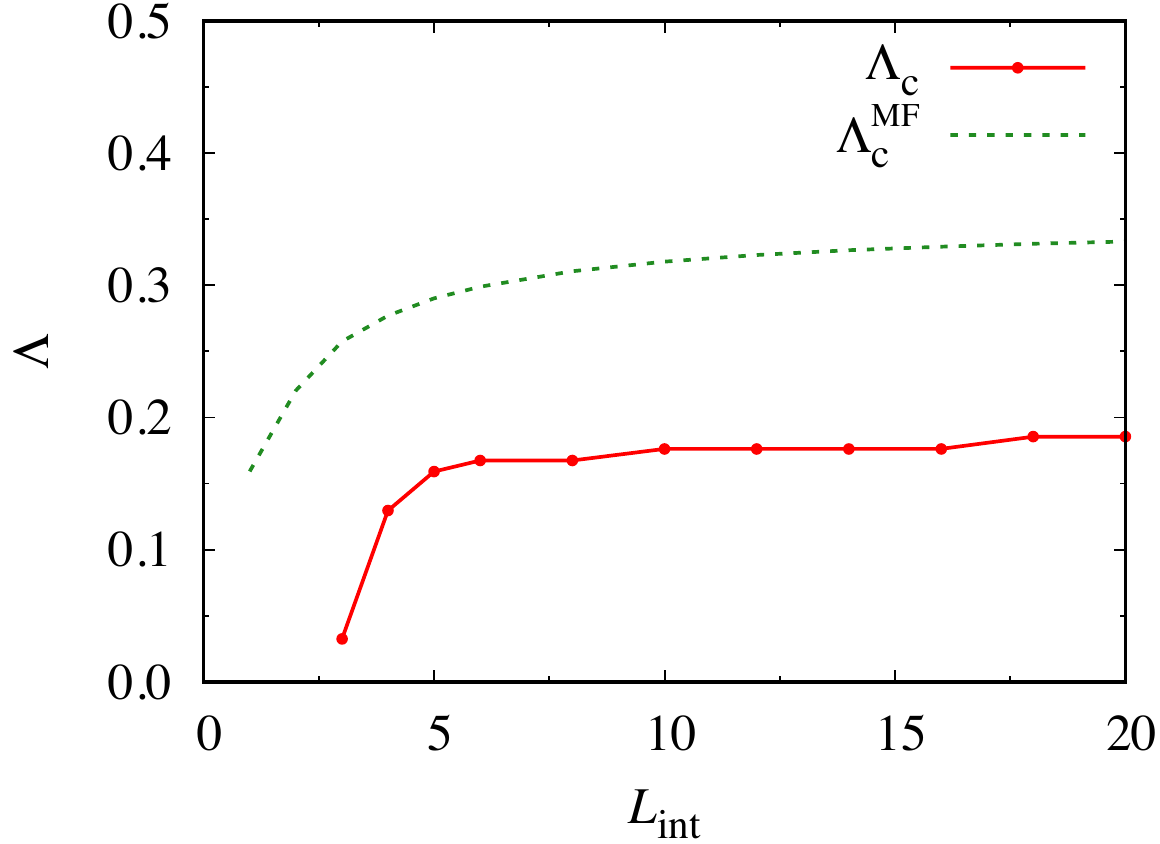}
    \caption{$L_{\mathrm{int}}$ dependences of $\Lambda_{\mathrm{c}}$ and $\Lambda_{\rm c}^{\rm MF}$ for the isotropic FM dipolar Kitaev model. }
    \label{fig:Lint_summary}
\end{figure}

Figure~\ref{fig:Lint_summary} summarizes the $L_{\mathrm{int}}$ dependences of $\Lambda_{\mathrm{c}}$ and $\Lambda_{\rm c}^{\rm MF}$, where $\Theta_{\rm CW}$ is calculated as in Sec.~\ref{subsec:frustration} within the range of $\|\mathbf{r}_{ij}\|_{\mathrm{b}}\leq L_{\mathrm{int}}$.
We find that $\Lambda_{\mathrm{c}}$ becomes nonzero for $L_{\mathrm{int}}\geq 3$ and rapidly increases to the saturation for $L_{\mathrm{int}}\gtrsim 6$. 
Meanwhile, $\Lambda_{\rm c}^{\rm MF}$ is nonzero for $L_{\mathrm{int}}\geq 1$ and increases rather gradually. The results indicate that further-neighbor interactions beyond second neighbors drastically reduce the degree of frustration, which results in the magnetic instability.


\section{Discussion}\label{discussion}
The effect of the long-range interaction on the spin liquid that we found for the dipolar Kitaev model contrasts with that for the dipolar Heisenberg model. The dipolar Heisenberg model is obtained as an implementation of the Heisenberg model with polar molecules trapped in an optical lattice, and has long-range Heisenberg-type interactions that are isotropic in spin space and decay in proportion to $r^{-3}$~\cite{Gorshkov2011a, Yan2013, Hazzard2014, Yao2018}.
Previous numerical studies showed that the ground states of the AFM dipolar Heisenberg models on square and triangular lattices can be quantum spin liquids, while the models with nearest-neighbor interactions stabilize long-range magnetic orders~\cite{Zou2017, Keles2018a, Keles2018}. 
Similar conclusions were drawn for the dipolar $XXZ$ models on the triangular lattice~\cite{Yao2018}. In stark contrast, in the dipolar Kitaev model studied in the present work, the spin liquid state realized by the nearest-neighbor Kitaev interaction is destabilized by the introduction of the long-range Kitaev-type interactions, and the ground state is replaced by a magnetically ordered state.

This difference can be attributed to different origins of the frustration. In the case of
the models with nearest-neighbor AFM Heisenberg interactions, the frustration is absent on the square lattice, and it is present but not strong enough to realize a quantum spin liquid state on the triangular lattice;
the frustration is enhanced by introducing long-range interactions as they compete with the nearest-neighbor one. In contrast, in the case of the Kitaev model, the frustration from the bond-dependent nearest-neighbor interactions is strong enough to stabilize the quantum spin liquid state with extremely short-range spin correlations~\cite{Kitaev2006,Baskaran2007}. In this case, the introduction of long-range interactions induce spin correlations between further neighbors. Our results indicate that the strong frustration from the nearest-neighbor interaction is relieved by the long-range interactions and the Kitaev spin liquid is replaced with magnetically ordered states.

Our results obtained by the PFFRG method conclude that it is difficult to realize the Kitaev quantum spin liquid by the implementation proposed for the ultracold polar molecules~\cite{Manmana2013,Gorshkov2013}. In the proposed setup, the long-range interactions inevitably appear because the magnetic interactions are implemented by the dipolar interactions between molecules. Hence, for the realization of the Kitaev quantum spin liquid, it is necessary to modify the long-range part of the interactions so that it does not hamper the spin liquid nature. Previous studies indicate that the Kitaev quantum spin liquid is fragile against the second-neighbor Kitaev interaction~\cite{Rousochatzakis2015}, while 
it remains stable for the Heisenberg interactions up to third-neighboring spins~\cite{Singh2012, Katukuri2014, Nishimoto2016}. Therefore, it may be possible to realize the Kitaev quantum spin liquid if one could replace the further-neighbor interactions of the Kitaev type with the Heisenberg type. 
In addition, it would be helpful to suppress the long-range part. Such an implementation in ultracold polar molecules is left for future studies.

\section{Summary}\label{conclusion}
To summarize, we have studied the ground state of a quantum spin model with long-range angle-dependent Kitaev-type interactions, which was proposed as an implementation of the Kitaev model in ultracold polar molecules, by using the PFFRG method. We clarified that, regardless of the spatial anisotropy of the interactions, the ground state is magnetically ordered in both FM and AFM cases: we found magnetic instabilities toward the FM and zigzag ordered states in the FM and AFM models, respectively. 
By calculation of the anisotropy parameter dependence of the critical cutoff scale, we concluded that the system is most frustrated and closest to the realization of the Kitaev quantum spin liquid when the interaction is isotropic in both cases. Our findings indicate that the quantum spin liquid ground state arising from the nearest-neighbor bond-dependent anisotropic interactions in the Kitaev model is destroyed by the long-range interactions. By varying the range of the interactions in the FM case, we elucidated that the Kitaev quantum spin liquid is unstable even for the third-neighbor interactions. 
Our results prompt a reconsideration of the implementation of the Kitaev-type interaction in polar molecules~\cite{Manmana2013, Gorshkov2013} to realize the Kitaev quantum spin liquid.
It would be helpful to suppress the long-range part or replace it by the Heisenberg-type.

\begin{acknowledgments}
K.F. thanks Yusuke Kato for constructive suggestions.
Parts of the numerical calculations have been done using the facilities of the Supercomputer Center, the Institute for Solid State Physics, the University of Tokyo, the Information Technology Center, the University of Tokyo, and the Center for Computational Material Science, Tohoku University.
This work was supported by Japan Society for the Promotion of Science (JSPS) KAKENHI Grant Nos. 19H05825 and 20H00122. K.F. was supported by the Program for Leading Graduate Schools (MERIT).
\end{acknowledgments}

\appendix
\section{Dependence on $\omega$ and $\Lambda$ grids}\label{appx:omega_Lambda}
\begin{figure}
    \centering
    \includegraphics[width=0.9\columnwidth, clip]{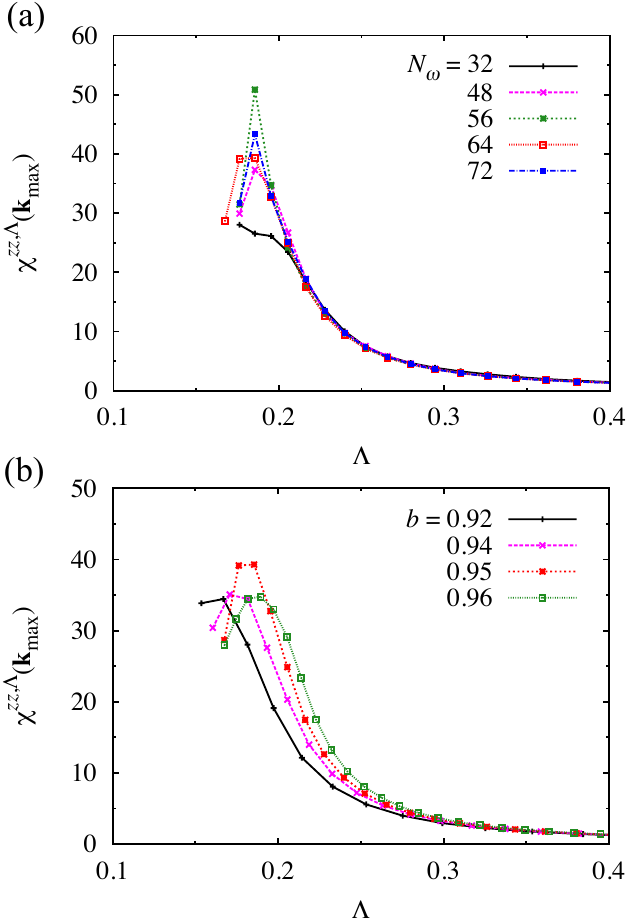}
    \caption{Spin susceptibility $\chi^{zz, \Lambda}(\mathbf{k}_{\mathrm{max}})$ for the isotropic FM dipolar Kitaev model as a function of $\Lambda$ while changing (a) the number of $\omega$ grids and (b) the multiplied factor $b$ to generate the $\Lambda$ grids.}
    \label{fig:Nw_b_suscep_flow}
\end{figure}

In this section, we discuss the effect of the discretization of $\omega$ and $\Lambda$ in the PFFRG calculations. Figure~\ref{fig:Nw_b_suscep_flow}(a) shows $\chi^{zz, \Lambda}(\mathbf{k}_{\rm max})$ for different $\omega$ grids, in the case of the isotropic FM dipolar Kitaev model ($\mathbf{k}_{\mathrm{max}}=\mathbf{0}$). 
Here, we discretize the frequency range of $10^{-4}\leq \omega\leq 250$ logarithmically with $N_\omega$ frequency points. The system size and the $\Lambda$ grids are the same as in the main text. We find that the data for $N_{\omega}\geq48$ show cusps at the same value of $\Lambda$ in the present resolution. Therefore, we conclude that $N_{\omega}=64$ is sufficiently large to estimate $\Lambda_{\rm c}$ and adopt it for the calculations in the main text. 

Meanwhile, Fig.~\ref{fig:Nw_b_suscep_flow}(b) shows $\chi^{zz, \Lambda}(\mathbf{k}_{\rm max})$ for different $\Lambda$ grids.  
Here, we discretize $\Lambda$ starting from $\Lambda_{\mathrm{max}}=500$ to $\Lambda_{\mathrm{min}}\simeq10^{-2}$ by multiplying the factor $b$ successively. The system size and the frequency grids are the same as in the main text. While $\chi^{zz, \Lambda}(\mathbf{k}_{\rm max})$ varies slightly while changing $b$, the data for $b\geq0.94$ show cusps at roughly the same $\Lambda$. Therefore, we adopt $b=0.95$ in the calculations in the main text. 

\section{System size dependence and 
finite-size scaling}\label{appx:scaling}
\begin{figure}
    \centering
    \includegraphics[width=0.9\columnwidth, clip]{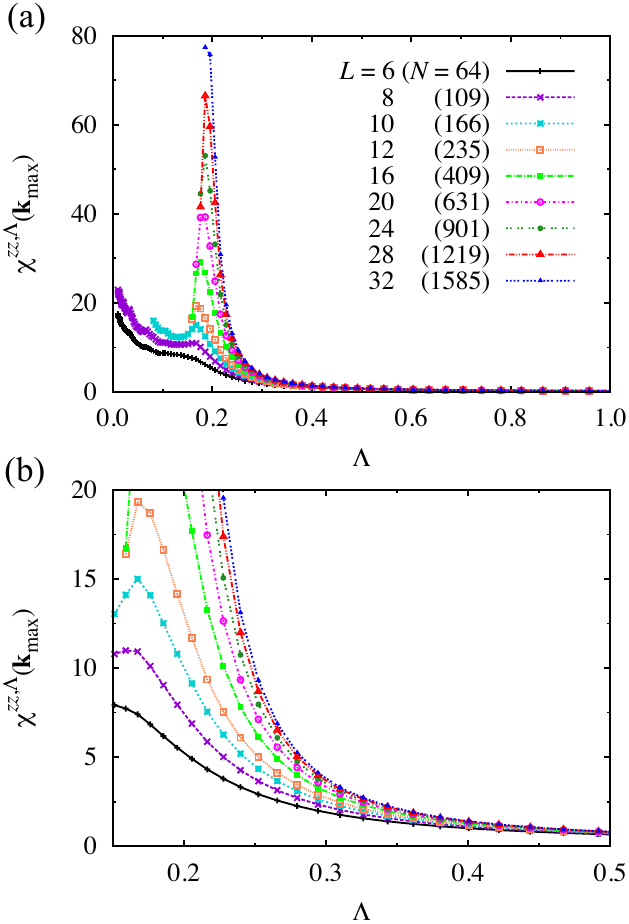}
    \caption{Spin susceptibility $\chi^{zz,\Lambda}(\bf{k}_{\mathrm{max}})$ for the isotropic FM dipolar Kitaev model as a function of $\Lambda$ for different system sizes. $L$ is the range of two-particle vertex functions, and $N$ is the corresponding number of sites. (b) is an enlarged figure of a part of (a).}
    \label{fig:L_suscep_flow}
\end{figure}

In this section, we discuss the system size dependence of the susceptibility. Figure~\ref{fig:L_suscep_flow} shows $\chi^{zz,\Lambda}(\bf{k}_{\mathrm{max}})$ for different system sizes, again for the isotropic FM dipolar Kitaev model ($\mathbf{k}_{\mathrm{max}}=\mathbf{0}$). The results indicate that $\chi^{zz,\Lambda}(\bf{k}_{\mathrm{max}})$ shows a divergent behavior as increasing the system size $L$, and that the value of $\Lambda_{\rm c}$ estimated from the cusp- or peak-like anomaly gradually becomes larger for larger $L$. 
To examine the behavior in the thermodynamic limit, from the analogy with the finite-size scaling at finite temperature, we assume the scaling relation as
\begin{equation}
    \frac{\chi^{zz,\Lambda}_{L}(\bf{k}_{\mathrm{max}})}{L^{2-\tilde{\eta}}}=g_{\Lambda}\left(\frac{\Lambda-\Lambda_{\mathrm{c}}^{\infty}}{\Lambda_{\mathrm{c}}^{\infty}}L^{1/\tilde{\nu}}\right),
\end{equation}
where $\chi^{zz,\Lambda}_{L}(\bf{k}_{\mathrm{max}})$ is the susceptibility for the system size $L$, $\tilde{\eta}$ and $\tilde{\nu}$ are ``critical exponents", $g_{\Lambda}$ is the scaling function, and $\Lambda_{\rm c}^{\infty}$ is the critical cutoff scale in the thermodynamic limit. 

\begin{figure}
    \centering
    \includegraphics[width=0.9\columnwidth, clip]{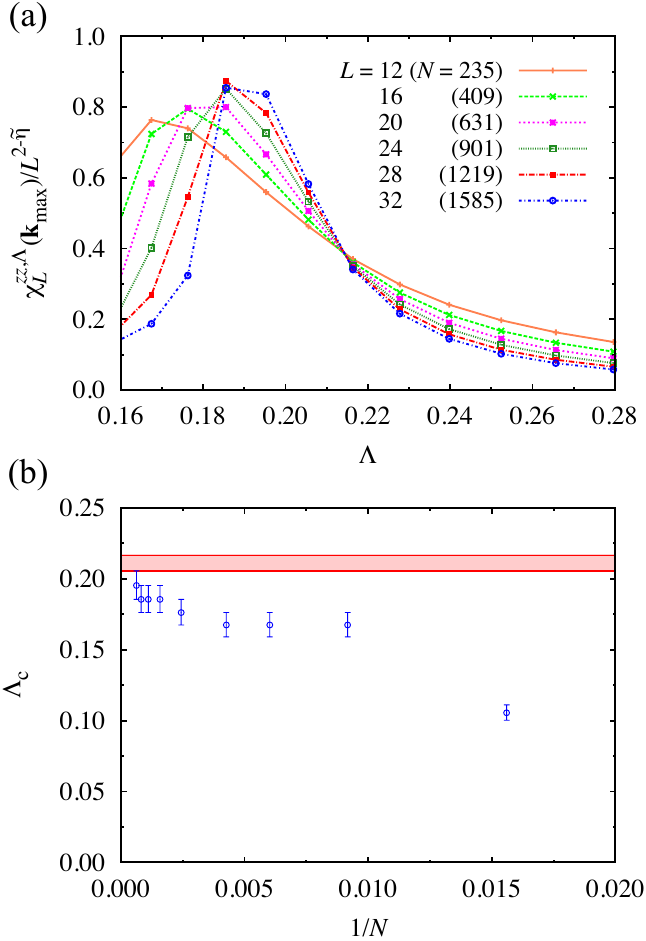}
    \caption{(a) $\Lambda$ dependence of the scaled susceptibility $\chi^{zz,\Lambda}_{L}(\mathbf{k}_{\mathrm{max}})/L^{2-\tilde{\eta}}$ for several values of $L$ and $\tilde{\eta}=0.7$. (b) System size dependence of $\Lambda_{\mathrm{c}}$ estimated from the cusps in Fig.~\ref{fig:L_suscep_flow}(a). The errorbars are given by the $\Lambda$ grids. The red hatched bar represents the estimate of $\Lambda_{\mathrm{c}}^{\infty}$ from the intersection in (a).}
    \label{fig:Lc_invN}
\end{figure}

We estimate the value of $\Lambda_{\rm c}^\infty$ by plotting $\chi^{zz,\Lambda}_{L}(\mathbf{k}_{\mathrm{max}})/L^{2-\tilde{\eta}}$ for different $L$ while changing $\tilde{\eta}$. 
We find that the data for $L\geq 12$ show an intersection for $\tilde{\eta}\simeq 0.7$; 
the result for $\tilde{\eta}=0.7$ is shown in Fig.~\ref{fig:Lc_invN}(a). 
From the intersection, we estimate that $\Lambda_{\rm c}^\infty$ is in the range of $0.206\lesssim\Lambda_{\mathrm{c}}\lesssim 0.216$. 
Figure~\ref{fig:Lc_invN}(b) shows the $N$ dependence of $\Lambda_{\rm c}$, together with the estimated range of $\Lambda_{\rm c}^\infty$, indicating that $\Lambda_{\rm c}$ converges slowly to $\Lambda_{\rm c}^\infty$. 
In the calculations in the main text, we adopt $L=20$ ($N=631$), for which $\Lambda_{\rm c}$ is underestimated roughly by $10$~\%.

\section{System size dependence of local susceptibility}\label{appx:local_suscep}
\begin{figure}
    \centering
    \includegraphics[width=0.9\columnwidth, clip]{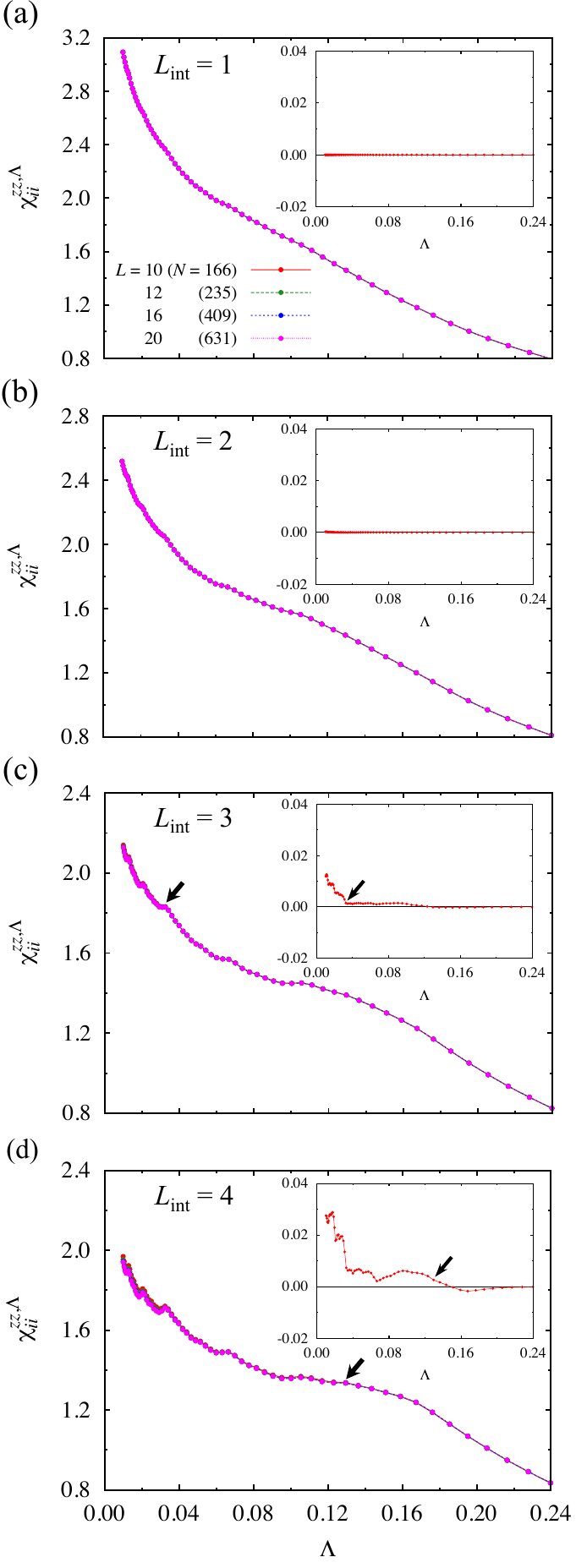}
    \caption{Local spin susceptibility $\chi^{zz, \Lambda}_{ii}$ as a function of $\Lambda$ for different system sizes with (a) $L_{\rm int}=1$, (b) $L_{\rm int}=2$, (c) $L_{\rm int}=3$, and (d) $L_{\rm int}=4$. The black arrows in (c) and (d) indicate the critical cutoff scale $\Lambda_{\mathrm{c}}$ determined in Fig.~\ref{fig:Lint_suscep_flow}. The insets show the differences between $L=10$ and $20$.}
    \label{fig:Lint_chi_ii}
\end{figure}

In this section, we investigate the system size dependence of the local spin susceptibility $\chi^{zz, \Lambda}_{ii}$ in the case of the isotropic FM dipolar Kitaev model. It was pointed out that $\chi^{zz, \Lambda}_{ii}$ is useful for detecting magnetic ordering since it shows a size dependence when the system becomes magnetically unstable~\cite{Kiese2019, Buessen2021}. 
Figure~\ref{fig:Lint_chi_ii} shows $\chi^{zz, \Lambda}_{ii}$ for different system sizes in the cases of $L_{\mathrm{int}} = 1$, $2$, $3$, and $4$. We find that the system size dependences for $L_{\rm int}=1$ and $2$ are negligibly small for all $\Lambda$, while the data for $L_{\rm int}=3$ and $4$ show system size dependences below $\Lambda_{\mathrm{c}}$ which are determined from the anomalies in $\chi^{zz,\Lambda}(\bf{k}_{\mathrm{max}})$ in Fig.~\ref{fig:Lint_suscep_flow}. 
These results not only confirm that our estimates of $\Lambda_{\mathrm{c}}$ are correct but also support our conclusion that the system shows a magnetic instability for $L_{\rm int}\geq 3$ in Sec.~\ref{subsec:long-range}.

\bibliography{library}

\end{document}